\documentclass[twocolumn]{aastex63}
\usepackage{amssymb}
\usepackage{array,multirow,hyperref}
\usepackage{comment}
\usepackage{enumerate}
\usepackage{bm}

\newcommand{\gaia}{{Gaia}}

\newcommand{\teff}{${T_{\rm{eff}}}$}

\newcommand{\Porb}{\ifmmode {P_{\rm orb}}\else${P_{\rm orb}}$\fi}

\newcommand{\Msun}{\ifmmode {{M_\odot}}\else{$M_\odot$}\fi}
\newcommand{\Mtot}{\ifmmode {{M_{\rm tot}}}\else{$M_{\rm tot}$}\fi}
\newcommand{\RV}{\ifmmode {{\rm RV}}\else RV \fi}
\newcommand{\bigG}{\ifmmode {\mathcal{G}}\else${\mathcal{G}}$\fi}

\submitjournal{ApJ}

\shorttitle{Theia 456}

\begin{document}

\title{{\bf A Young, Low-Density Stellar Stream in the Milky Way Disk: Theia 456}}

\newcommand{\amnh}{American Museum of Natural History, 200 Central Park West, New York, NY 10024, USA}
\newcommand{\columbia}{Department of Astronomy, Columbia University, 550 West 120th Street, New York, NY 10027, USA}
\newcommand{\westernwashington}{Department of Physics \& Astronomy, Western Washington University, Bellingham, WA 98225-9164 USA}

\correspondingauthor{Jeff J.~Andrews}
\email{jeffrey.andrews@northwestern.edu}

\author[0000-0001-5261-3923]{Jeff J.~Andrews}
\affiliation{Center for Interdisciplinary Exploration and Research in Astrophysics (CIERA), 
1800 Sherman Ave., 
Evanston, IL 60201, USA}

\author[0000-0002-2792-134X]{Jason L.~Curtis}
\affiliation{\columbia}
\affiliation{\amnh}

\author[0000-0003-2481-4546]{Julio Chanam{\'e}}
\affiliation{Instituto de Astrofísica, Pontificia Universidad Cat{\'o}lica de Chile, Av. Vicu{\~n}a Mackenna 4860, 782-0436 Macul, Santiago, Chile}

\author[0000-0001-7077-3664]{Marcel A.~Ag\"{u}eros}
\affiliation{\columbia}

\author[0000-0001-7203-8014]{Simon C.~Schuler}
\affiliation{University of Tampa, Department of Chemistry, Biochemistry, and Physics, Tampa, FL 33606, USA}

\author[0000-0002-5365-1267]{Marina Kounkel}
\affiliation{\westernwashington}

\author[0000-0001-6914-7797]{Kevin~R.~Covey}
\affiliation{\westernwashington}

\begin{abstract}
Our view of the variety of stellar structures pervading the local Milky Way has been transformed by the application of clustering algorithms to the \gaia\ catalog. In particular, several stellar streams have been recently discovered that are comprised of hundreds to thousands of stars and span several hundred parsecs. We analyze one such structure, Theia 456, a low-density stellar stream extending nearly 200~pc and 20$^{\circ}$ across the sky. By supplementing \gaia\ astrometric data with spectroscopic metallicities from LAMOST and photometric rotation periods from the Zwicky Transient Facility (ZTF) and the Transiting Exoplanet Survey Satellite (TESS), we establish Theia 456's radial velocity coherence, and we find strong evidence that members of Theia 456 have a common age ($\simeq$175 Myr), common dynamical origin, and formed from chemically homogeneous pre-stellar material ([Fe/H] = $-$0.07~dex). Unlike well-known stellar streams in the Milky Way, which are in its halo, Theia 456 is firmly part of the thin disk. If our conclusions about Theia 456 can be applied to even a small fraction of the remaining $\simeq$8300 independent structures in the Theia catalog, such low-density stellar streams may be ubiquitous. We comment on the implications this has for the nature of star-formation throughout the Galaxy.
\end{abstract}


\section{Introduction}
\label{sec:intro}

It is generally believed that most stars form in stellar clusters embedded within giant molecular clouds \citep[cf.~review in][]{lada2003}. However, many of these clusters are not massive enough to be bound, and only a minority of stars are formed within bound clusters, with the specific fraction dependent upon the definition of cluster used \citep{bressert2010}. Some fraction of these stellar clusters may then survive for millions or even billions of years, as stellar associations, moving groups, and open clusters, but ultimately most stars disperse into the Milky Way field as single stars \citep{fall2005, goodwin2006}. 

Studies of stellar populations in the Galaxy provide important constraints on the details of this scenario. These studies have been revolutionized by the \gaia\ catalog, with its vastly improved astrometric precision and nearly two billion separately identified objects in its Early Data Release 3 \citep[EDR3;][]{gaia_EDR3_summary, gaia_EDR3_astrometry,gaia_EDR3_photometry}. In particular, researchers have uncovered new details about the population of Milky Way globular clusters \citep[e.g.,][]{vitral2021}, open clusters \citep[e.g.,][]{castro-ginard2020}, moving groups and young stellar associations \citep[e.g.,][]{ujjwal2020} stellar streams \citep[e.g.,][]{helmi2020}, and the origin of the Milky Way halo \citep[e.g.,][]{belokurov2018, helmi2018}.

Some of the most exciting discoveries have been produced by the application of modern statistical and machine learning algorithms to the \gaia\ data. For instance, using a combination of a wavelet decomposition technique and the unsupervised clustering algorithm DBSCAN, \citet{meingast2019} discovered the Pisces--Eridanus/Meingast~1 stream, a nearly 400-pc-long structure comprised of 256 stars. Further studies by \citet{ratzenbock2020} and \citet{roser2020}  used support-vector machine and convergent point methods, respectively, to expand the number of member stars in this stream to over 10$^3$.

Other recent, similar  discoveries include that of \citet{tian2020}, who identified a young (30--40~Myr), 200~pc long and 80~pc wide, stellar stream comprised of several thousand members.  And \citet{beccari2020} detected a 260-pc-long filamentary structure that links two previously known clusters, BBJ~1 and NGC~2547. 

These discoveries may only be the tip of the iceberg. \citet{kounkel2019} and \citet{kounkel2020} applied HDBSCAN \citep{campello2013, mcinnes2017}, a variation of DBSCAN, to the \gaia\ data release 2 \citep[DR2;][]{Gaia_DR2_summary, Gaia_astrometry}. These authors' Theia catalog, which extends to a distance of 3~kpc from the Sun, includes 8292 separate stellar structures, many of which are previously undetected, large, low-density filaments or streams, comprised of hundreds to thousands of stars and spanning hundreds of pc.

Whereas previously known stellar streams were all found within the Milky Way halo \citep[e.g., Pal~5;][]{odenkirchen2001} and therefore the likely result of the tidal disruption of infalling satellite galaxies, these recently discovered  structures exist within the Milky Way disk. As such, their origin is  unclear. Both simulations \citep{renaud2013} and observations \citep{van_loon2003} show that Galactic resonances produce overdensities in the Milky Way disk; these are collections of unassociated stars that happen to be caught within the same resonance. Alternatively, low-density stellar structures could have been formed from the same pre-stellar material, and we are observing them in the midst of dissipation \citep{kamdar2019}. 

Two follow-up studies of Pisces--Eridanus/Meingast~1 compared its properties to that of the benchmark Pleiades open cluster. \citet{curtis2019} used rotation period measurements for stars in both structures to demonstrate that a large fraction of Pisces--Eridanus/Meingast~1 stars share a common age of $\approx$120~Myr, the age of the Pleiades. \citet{hawkins2020} found that its stars have similar metallicity, and based on Li abundances,  age, reinforcing the claim that they have a common origin. Follow-up observations such as these are critical to testing whether stars in these new structures share a common age and chemical origin, and therefore in elucidating their origin.

In this work, we assess the origin of another exciting, young, low-density stellar structure within the Milky Way disk, Theia 456. In Section~\ref{sec:characteristics}, we demonstrate that after culling outliers, Theia 456 members have consistent kinematics, metallicity, and age. We analyze the dynamical origin of Theia 456 in Section~\ref{sec:dynamics} by integrating orbits in a Milky Way potential. We discuss the implication our results have for star formation and Milky Way substructure in Section~\ref{sec:conclusions}, and provide some concluding thoughts.

\section{Characteristics of Theia 456}
\label{sec:characteristics}

To understand the structure and characteristics of Theia 456, we first analyze the \gaia\ EDR3 kinematics of candidate members identified by \citet{kounkel2019} in the \gaia\ DR2 catalog. After removing outliers, we provide an updated catalog of Theia 456 stars. We then cross-match the positions of these members with several additional catalogs: the Large Sky Area Multi-Object Fiber Spectroscopic Telescope (LAMOST) DR5 spectroscopic catalog \citep{LAMOST, LAMOST_DR1}, and the Transient Exoplanet Survey Satellite \citep[TESS;][]{TESS}, and Zwicky Transient Facility \citep[ZTF;][]{ZTF_overview, ZTF_data} photometric databases.  

Combined with astrometric, radial velocity, and photometric data from \gaia, these catalogs provide a complementary perspective which allows us to analyze the kinematic characteristics, metallicity, and age of Theia 456, each of which we describe in turn below.

\subsection{Theia 456 Membership}
\label{sec:membership}

\begin{figure}
    \begin{center}
    \includegraphics[trim=0.2cm 0.1cm 0.2cm 0.2cm, clip=True, width=1.0\columnwidth]{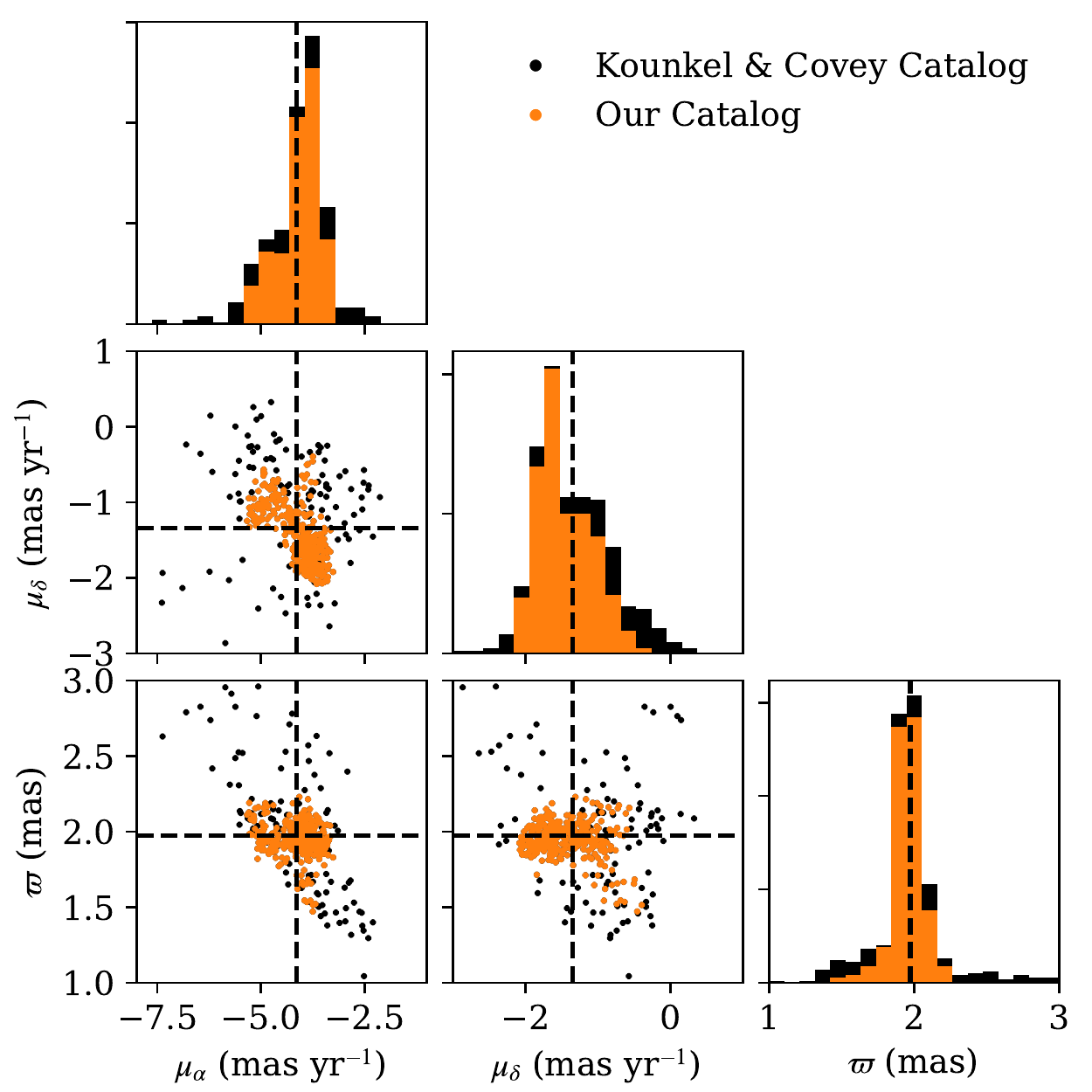}
    \caption{\gaia\ EDR3 astrometry for stars in the initial \cite{kounkel2019} catalog of candidate Theia 456 members (black) and our updated catalog (orange), in which many of the outliers have been removed. The stars in the updated catalog have consistent astrometric characteristics. Dashed lines indicate the median for each parameter. Some remaining scatter may be due to measurement errors contamination, and the combination of perspective effects and differences in the Milky Way's potential varying over the length of Theia 456.}  
    \label{fig:astrometry}
    \end{center}
\end{figure}

We first start by updating the astrometric characteristics of the 468 reported members of Theia 456 identified by \citet{kounkel2019} to the latest \gaia\ EDR3 astrometric catalog from the original DR2 catalog. While the astrometric measurement precision for these stars improved between the two catalogs, we found the DR2 data to be quite accurate. In Figure \ref{fig:astrometry} we provide a corner plot showing the EDR3 astrometric characteristics (proper motion Right Ascension, $\mu_{\alpha}$, and Declination, $\mu_{\delta}$, as well as parallaxes, $\varpi$) of the 468 candidate members identified by \citet{kounkel2019} as black points and histograms. With a typical precision $\lesssim$0.1 mas in parallax and $\lesssim$0.1 mas yr$^{-1}$ in proper motion, the scatter in the black points is larger than can be accounted for by measurement uncertainties, suggesting the initial catalog contains some contaminating stars.

\begin{figure}
    \begin{center}
    \centerline{\includegraphics[width=1.0\columnwidth, trim=0.4cm 0.4cm 0.3cm 0.3cm, clip=True]{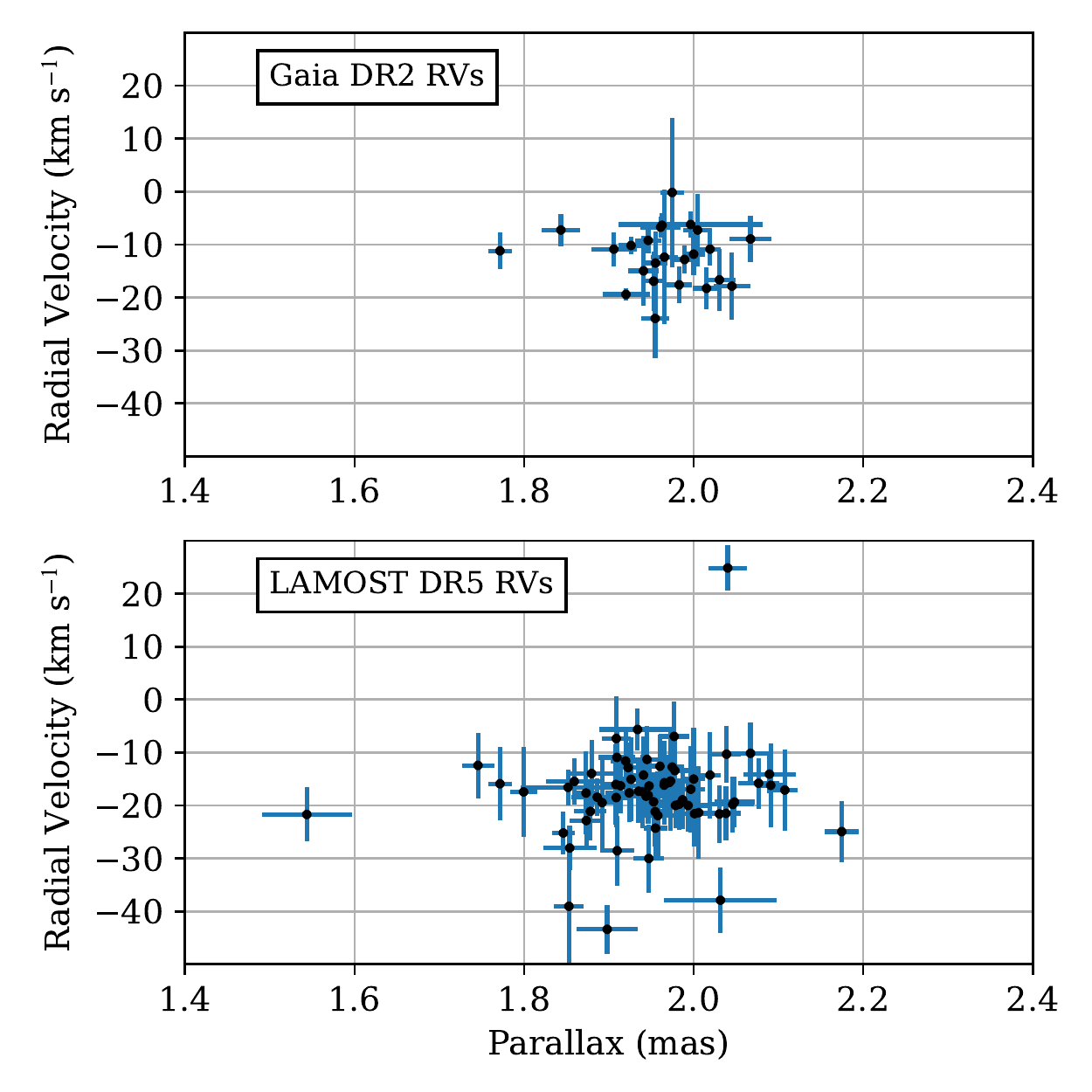}}
    \caption{The radial velocities of Theia 456 stars as measured by either \gaia\ (top panel) or LAMOST (bottom panel) as a function of their \gaia\ parallaxes. The stars in Theia 456 were identified by their astrometric data only; the overall consistency of the radial velocities provides confirmation that Theia 456 is a kinematically coherent structure.}  
    \label{fig:RV}
    \end{center}
\end{figure}

\begin{figure}
    \begin{center}
    \centerline{\includegraphics[width=0.8\columnwidth, trim=0.3cm 0.4cm 0.3cm 0.3cm, clip=True]{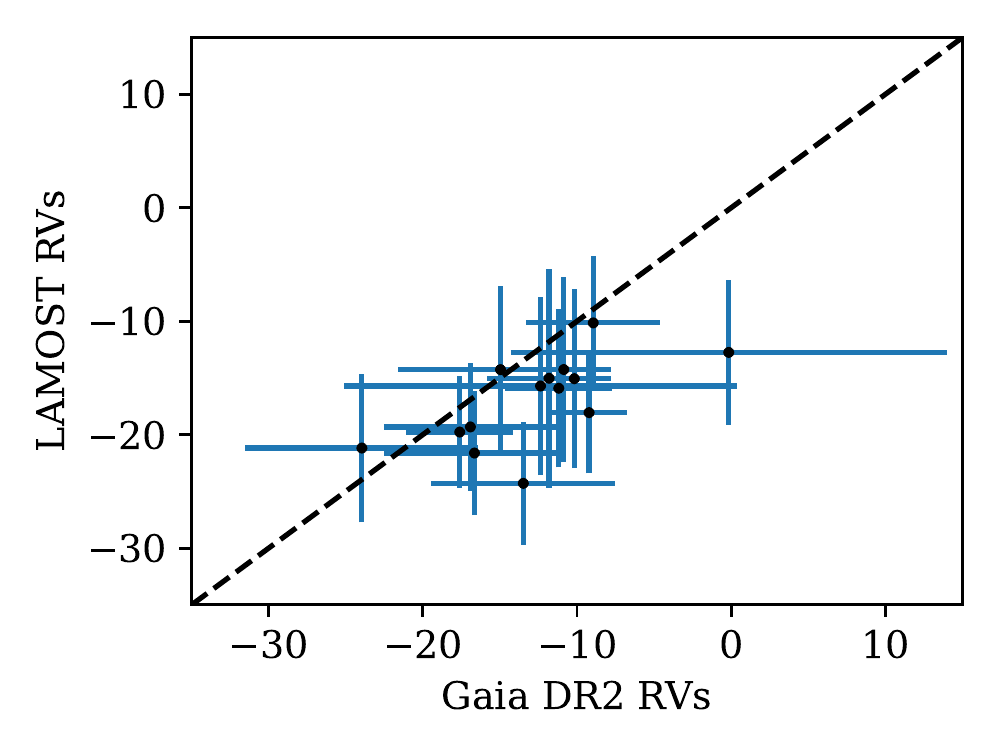}}
    \caption{We compare the radial velocities for the 14 stars in Theia 456 which have measurements in both \gaia\ DR2 and LAMOST. The radial velocities are generally consistent between the two experiments.}  
    \label{fig:RV_compare}
    \end{center}
\end{figure}

To refine our catalog of Theia 456 members, we applied the unsupervised clustering algorithm DBSCAN to the three-dimensional astrometric data: $\mu_{\alpha}$, $\mu_{\delta}$, and $\varpi$. We use the implementation provided by {\tt scikit-learn} \citep{scikit-learn}, with an {\tt eps} value of 0.3, resulting in a refined catalog of 362 members. Although we found a value of 0.3 to be a good compromise between catalog size and purity, the resulting membership catalog depends quite strongly on the {\tt eps} value chosen. Our catalog may very well continue to contain some contamination, and it is almost certainly incomplete. 
In deriving their catalog \citet{kounkel2019} made relatively stringent quality cuts, selecting stars with $\varpi / \sigma_{\varpi} > 10$ and $\sigma_{\rm G} \lesssim0.03$ mag, among other constraints. Even within the \gaia\ catalog there may be more as-yet unidentified Theia 456 members.
We show as orange points in Figure \ref{fig:astrometry} the astrometric characteristics of our updated catalog of Theia 456 members; the most significant outliers are removed from this updated sample. We use our updated membership catalog (which we provide in Appendix \ref{a:rotation}) throughout the remainder of this work.

\subsection{Kinematic Characteristics}
\label{sec:kinematic_characteristics}

\begin{figure*}
    \begin{center}
    \includegraphics[width=0.8\textwidth]{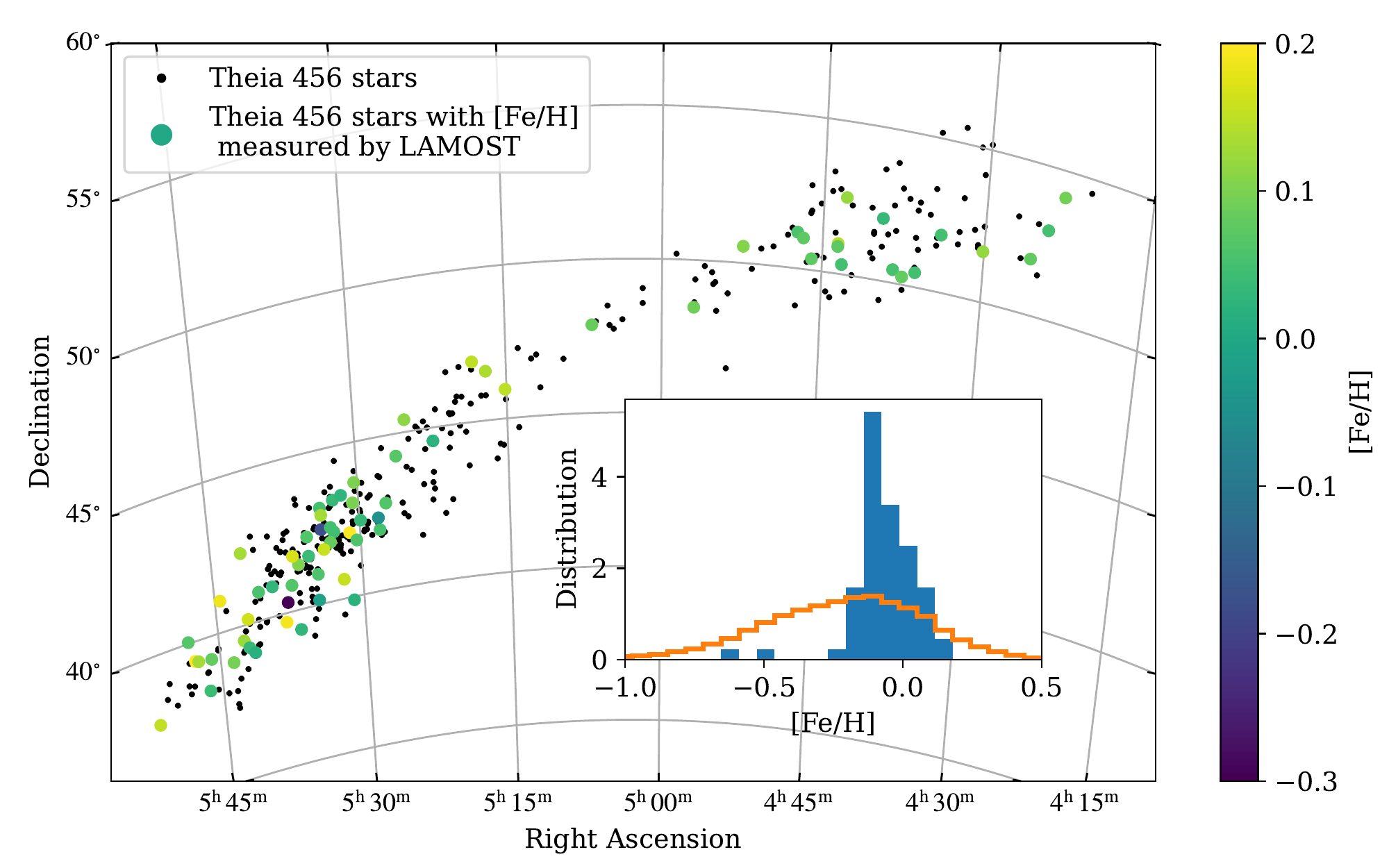}    
    \caption{Positions of the 69 stars in Theia 456 with [Fe/H] measured by LAMOST (colored markers). Black points indicate the positions of the remaining 293 stars in Theia 456, as identified by \citet{kounkel2019}. The blue histogram in the inset shows the [Fe/H] distribution. With the exception of a few outliers, the stars in Theia 456 are tightly distributed around [Fe/H]~$=-0.07$~dex. Compared to the orange histogram, which shows the distribution of [Fe/H] of all LAMOST stars within the same field of view, Theia 456 stars are sharply concentrated. }  
    \label{fig:Fe_H}
    \end{center}
\end{figure*}

Taking only the orange points and distributions in Figure \ref{fig:astrometry}, we find that the stars in our updated Theia 456 membership catalog show a clear overdensity, with median astrometric values: $\mu_{\alpha}\simeq -$4.1$\pm0.5$~mas~yr$^{-1}$, $\mu_{\delta}\simeq-1.5\pm0.4$~mas~yr$^{-1}$, and $\varpi\simeq1.9\pm0.1$~mas. Although the distribution extends slightly beyond the typical astrometric measurement precision of \gaia\ EDR3 ($\sim$0.1~mas yr$^{-1}$ in proper motion and $\sim$0.02~mas for parallax), some variation is expected for an extended distribution of stars; Theia 456 subtends a large enough angle on the sky that differences in our viewing perspective as well as in the Galactic potential across different ends of the structure can become relevant.

\gaia\ DR2 included radial velocities for seven million stars with 3800~K$<$\teff$<$~7000~K \citep{DR2RadialVelocities}, measured using the infrared \ion{Ca}{2} triplet. For Theia 456 members, these typically have a measurement precision of a few km s$^{-1}$, depending on the star's apparent magnitude, stellar type, rotational velocity, and multiplicity. At the same time, the spectroscopic catalog LAMOST has measured radial velocities for a subset of Theia 456 stars, with a precision comparable to that of \gaia. We cross-match our catalog of Theia 456 stars with LAMOST, finding 69 stars with positions in both catalogs matching to within 0.5$^{\prime\prime}$.

In Figure~\ref{fig:RV}, we show these radial velocities for members of Theia 456 as a function of the stars' \gaia\ parallax. The \gaia\ and LAMOST data (24 and 69 stars, respectively) both show a cluster of stars with similar radial velocities (\gaia: $-11.5\pm$5.3 km s$^{-1}$; LAMOST: $-17.3\pm$8.4 km s$^{-1}$). These are comparable to the median radial velocity measurement precision for Theia 456 members of 5 km s$^{-1}$ for \gaia\ and 6 km s$^{-1}$ for LAMOST, suggesting that current radial velocity measurements are not precise enough to reliably detect any internal velocity dispersion. A small subset of 14 stars were observed by both instruments, and we compare their respective radial velocity measurements in Figure \ref{fig:RV_compare}. The two independent radial velocity measurements for these 14 stars are consistent within measurement uncertainties. We note that there appears to be a systematic difference between measurements from the two experiments, consistent with the $-5$\,km\,s$^{-1}$ offset for LAMOST radial velocities reported by \citet{anguiano2018} when comparing against radial velocities from APOGEE \citep{majewski2017}. Whatever offset exists is smaller than typical measurement uncertainties.

The error bars in Figure \ref{fig:RV} indicate that, with the exception of a few discrepant points, much of this scatter is likely due to measurement uncertainties. We note that in their algorithm, \citet{kounkel2019} did not use radial velocities to identify stellar structures. Furthermore, we did not use radial velocities in producing our refined membership catalog in Section \ref{sec:membership}. The overall radial velocity consistency of stars in Figure~\ref{fig:RV} serves as a confirmation of the kinematic coherence of Theia 456.

\subsection{Metallicity}
\label{sec:metallicity}

\begin{figure}
    \begin{center}
    \centerline{\includegraphics[trim=0.3cm 0.5cm 0.2cm 0.3cm, clip=True,
    width=1\columnwidth]{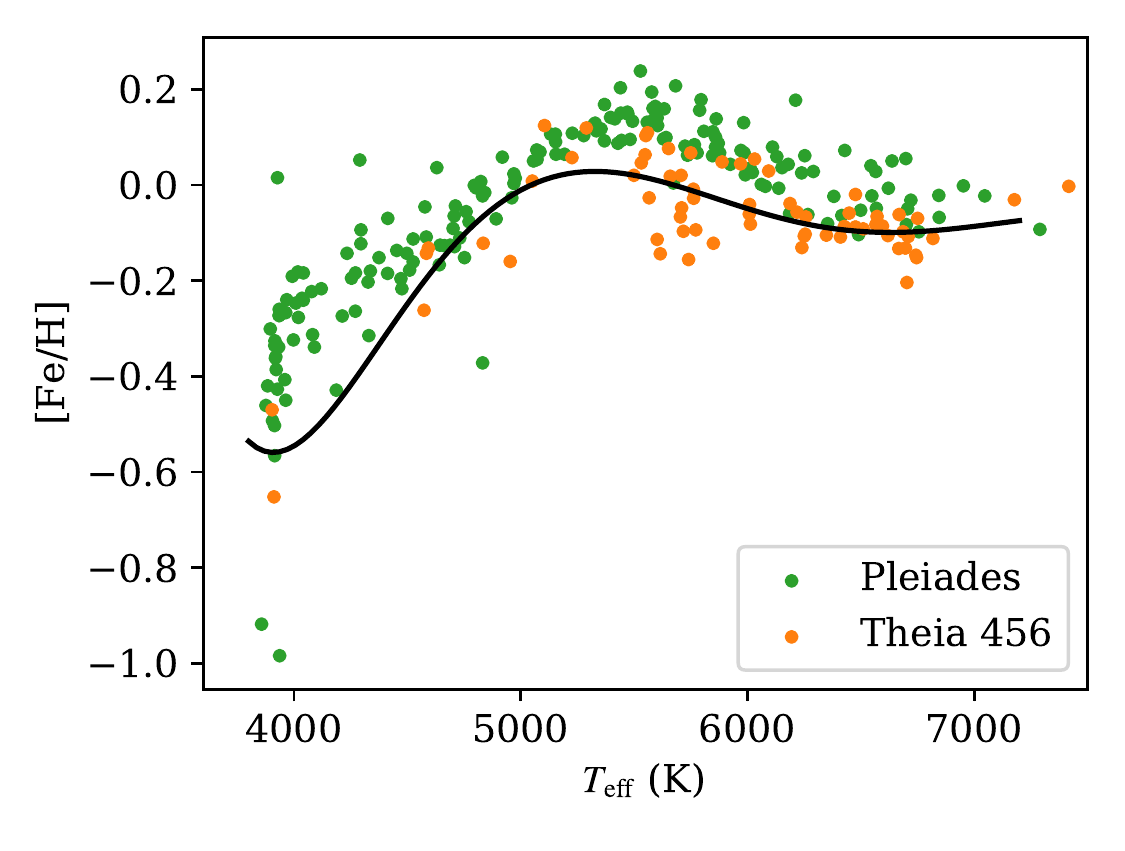}}    
    \caption{ The LAMOST [Fe/H] of Pleiads (green) and Theia 456 stars (orange) as a function of the LAMOST-measured $T_{\rm eff}$. The well-studied Pleiades has been previously shown using high-resolution spectroscopy to have a mean [Fe/H]$=+0.01$, with a star-to-star scatter $<$0.11 dex \citep{schuler2010}. Where we would expect to find a consistent [Fe/H] throughout the cluster, Pleiads exhibit a systematic trend in LAMOST-measured [Fe/H] with respect to $T_{\rm eff}$, a feature also seen for Theia 456 stars. We also show a seventh order polynomial fit (black line) to Theia 456 stars. Comparing the polynomial fit to our Theia 456 stars gives us an updated dispersion in [Fe/H] of 0.07 dex, narrower than the inset to Figure \ref{fig:Fe_H} suggests.} 
    \label{fig:Fe_H_Teff}
    \end{center}
\end{figure}

Figure \ref{fig:Fe_H} shows the positions of all the stars in Theia 456, highlighting the subset that have been observed by LAMOST. The LAMOST targets, which are spread out across the length of the stream, tend to have a near-Solar metallicity: their median metallicity is [Fe/H]~$=-0.07$~dex and a standard deviation of 0.12~dex, with the full distribution shown as a blue histogram in the inset of Figure~\ref{fig:Fe_H}.

Typical statistical errors in [Fe/H] for LAMOST stars are $\lesssim$0.02 dex, much smaller than the spread observed in Theia 456 stars. However, repeat measurements by LAMOST of the same stars indicate a somewhat larger uncertainty of $\simeq$0.06 dex \citep{LAMOST_DR1}. By comparing the measured metallicities of stars in widely separated stellar binaries, \citet{andrews2018} found LAMOST measurements of [Fe/H] to be consistent with this somewhat larger measurement precision. For comparison, we note that recent analysis of the Pisces--Eridanus/Meingast 1 stream by \citet{hawkins2020} indicates a [Fe/H] spread of 0.07 dex (most of which is driven by measurement uncertainties), suggesting that the inherent spread one might expect in such a structure should be similar to or smaller than the precision of the LAMOST metallicity measurements. 

To look more deeply into LAMOST's metallicity measurement accuracy, we cross-match the \gaia\ DR2 sample of Pleiades stars with LAMOST, finding 184 Pleiads with [Fe/H] measured by LAMOST. In Figure \ref{fig:Fe_H_Teff} we compare the \teff\ with the [Fe/H] for these 184 stars as green markers. While previous analysis using high-resolution spectroscopy \citep[e.g.,][]{schuler2010} has shown that the Pleiades is co-chemical, with a star-to-star scatter in [Fe/H] $<$0.11 dex, Figure \ref{fig:Fe_H_Teff} shows a clear systematic dependence in LAMOST data of [Fe/H] on \teff. Figure \ref{fig:Fe_H_Teff} additionally shows these quantities for Theia 456 stars as orange markers. There is a slight offset in the two data sets, consistent with Theia 456 having a lower metallicity. Critically, the dependence of [Fe/H] on \teff\ suggests that the broad distribution of [Fe/H] of Theia 456 members (at least compared to the typical measurement uncertainty) is due to a systematic effect in LAMOST [Fe/H]. As a clear example, the two Theia 456 stars with \teff$\simeq4000$~K both have [Fe/H]$<-0.4$ dex, corresponding to the low metallicity outliers in the inset of Figure \ref{fig:Fe_H}. 

To derive an improved [Fe/H] dispersion, we fit the $T_{\rm eff}$ and [Fe/H] for Theia 456 stars to with a seventh order polynomial, which we display in Figure~\ref{fig:Fe_H_Teff} as a black line. Comparison between the polynomial fit and Theia 456 stars provides an updated, de-trended estimate of the [Fe/H] dispersion of 0.07 dex, consistent with the magnitude of systematic uncertainties in LAMOST-measured [Fe/H] as well as the dispersion previously reported for the Pisces--Eridanus/Meingast 1 stream by \citet{hawkins2020}. Combined with the kinematic coherence described in Section \ref{sec:kinematic_characteristics}, we therefore conclude that Theia 456 stars likely have a common origin.

%We therefore conclude that the metallicity distribution of Theia 456 stars is even tighter than the inset to Figure \ref{fig:Fe_H} suggests. Moreover, combined with the kinematic coherence described in Section \ref{sec:kinematic_characteristics}, we conclude that Theia 456 stars likely have a common origin.

\begin{figure}
    \begin{center}
    \includegraphics[width=1.0\columnwidth]{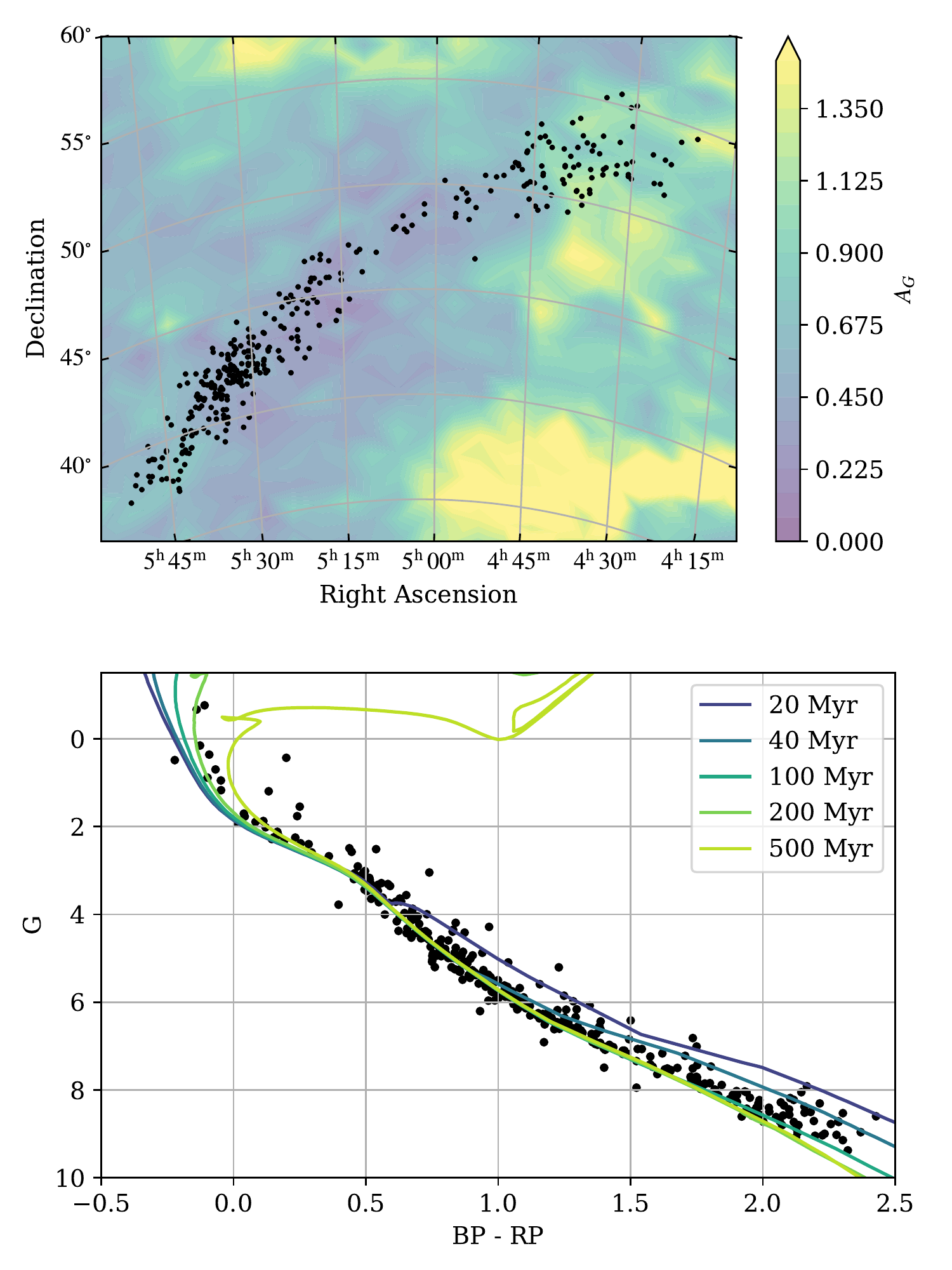}
    \caption{{\it Top panel---}Positions of Theia 456 stars (black points) overlaid on a dust extinction map compiled from stars at the same distance for which \gaia\ has measured the $A_G$. The extinction varies by over a magnitude across the extent of Theia 456. {\it Bottom panel---} Extinction-corrected color--magnitude diagram for Theia 456 stars. The scatter in the CMD can be due to a combination of effects, including the presence of binaries, contaminating stars, and incorrect extinction estimation. Comparison with isochrones from PARSEC (colored lines) suggest an age of $\simeq$200 Myr.}
    \label{fig:CMD}
    \end{center}
\end{figure}

\subsection{The challenges of obtaining a robust isochrone age  for Theia 456}
\label{sec:isochrone}

Isochrone fitting, although not the only approach, has traditionally been the preferred method for dating stellar populations \citep[for a review, see][]{soderblom2010}. In their catalog \citet{kounkel2019} provide an age of $\simeq$165 Myr for Theia 456, derived from a combination of a bespoke convolutional neural network and a Bayesian Markov chain Monte Carlo method \citep{von_hippel2006}. Despite their complexity, these methods are both based on stellar multi-band photometry, a form of isochrone fitting. \citet{kounkel2019} suggest their technique provides an age accuracy of 0.15 dex, which leads to an age range of 115--235 Myr for Theia 456.

Accurate isochrone fitting requires knowledge of the interstellar reddening along the line-of-sight, as fixing the metallicity and reddening is key to obtaining a robust isochrone age. Typically, stellar clusters are either nearby or subtend a small solid angle on the sky, so  extinction is minimal or at least constant across the population. Stellar streams present much more of a challenge, as one may need to estimate extinction for each star. Theia 456 spans nearly 25 degrees across the sky and is both sufficiently distant ($\simeq$500 pc) and sufficiently near the Galactic Plane that the amount of intervening dust could be consequential and vary significantly from star to star. It is therefore worthwhile to revisit the age of Theia 456.

To account for extinction, we generate our own \gaia\-based dust map for Theia 456. We select the subset of stars in the field of Theia 456 that have a parallax in the range (1.85, 2.05) mas and for which \gaia\ has measured a $G$-band extinction $A_G$, and perform a $k$-nearest neighbors interpolation scheme where $k=10$\footnote{This parallax range is chosen based on the median Theia 456 member parallax of 1.95 mas.}. The top panel of Figure \ref{fig:CMD} shows the resulting $A_G$ map and confirms that Theia 456 is a challenging case: the extinction for stars in Theia 456 varies by over a magnitude across its span.

In the bottom panel of Figure \ref{fig:CMD} we show the resulting color--magnitude diagram (CMD), where we use our dust map to separately account for each star's extinction and reddening. We used a constant distance of 512 pc in calculating the absolute $G$ magnitudes.\footnote{This distance is taken from the inverse of the median parallax of Theia 456 stars. While it is unlikely that the structure of Theia 456 lies exactly perpendicular to the Sun, we have found no significant variation in the distances to stars over its extent (see Section \ref{sec:gradient}).} The CMD shows some scatter, more than what is seen in nearby star clusters like the Pleiades or Hyades. Several factors could be contributing to this effect, including contamination, the presence of stellar binaries and blue stragglers, and an imperfect extinction map \citep{soderblom2010}. Nevertheless, near the turn-off most stars lie close to, or a bit red-ward of, the 200 Myr isochrone \citep[taken from the PARSEC models][]{bressan2012}, consistent with the 165$^{+70}_{-50}$ Myr age estimate provided by \citet{kounkel2019}. 

\begin{figure}
    \begin{center}
    \includegraphics[width=1.0\columnwidth]{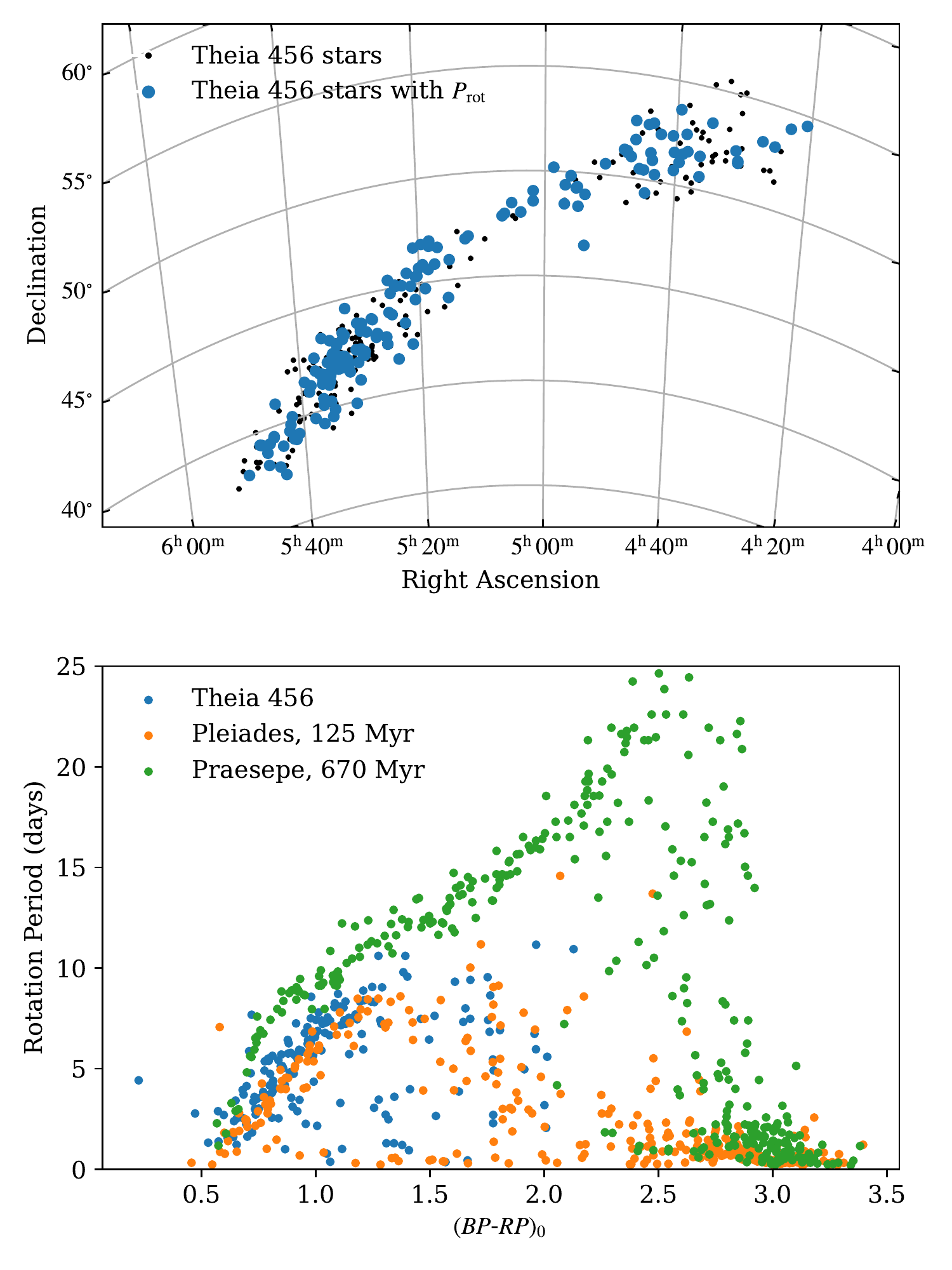}
    \caption{{\it Top panel---}Theia 456 stars for which we measure a rotation period from either a TESS or a ZTF lightcurve. These stars span the extent of Theia 456. {\it Bottom panel---}\gaia\ color--period diagram comparing the distribution for Theia 456 (blue markers) to that of stars from the Pleiades (orange) and Praesepe (green). Theia 456 lies between the Pleiades and Praesepe, suggesting an age between the two. }
    \label{fig:CPD}
    \end{center}
\end{figure}

\subsection{A rotation-based age}
\label{sec:rotation}

Rotation periods provide an additional, robust estimator for stellar ages that is independent of isochrone fitting \citep{barnes2007, mamajek2008}. Magnetic braking causes stars of different masses to slow their rotation at different rates. By comparing the photometric colors of the stars in a stellar population with their rotation periods, a clear sequence can be used to determine the stars' ages. Rather than match theoretical models for stellar spin-down, constraints can be made by comparing to the corresponding distributions in stellar clusters with known ages. Any reasonable comparison requires a large enough sample of stars with high-cadence photometric observations like that provided by surveys from TESS and ZTF. 

To obtain this sample, we cross-match the stars in Theia 456 with both photometric surveys, finding 132 and 164 stars have well-sampled photometric data from ZTF and TESS, respectively, with 80 stars sampled by both experiments. Our analysis of these light curves follows that described in detail elsewhere \citep[e.g.,][]{curtis2020}. Briefly, we use an automated pipeline that applies a Lomb-Scargle periodogram to these light curves to derive rotation periods. After visual inspection to confirm the quality of these periodicities, we derive robust rotation periods for 216 stars in Theia 456 (see Appendix~\ref{a:rotation} for details). For stars with rotation periods measured by both experiments, we use the average.

Crucially, these stars span the full stream (see top panel, Figure~\ref{fig:CPD}).
The bottom panel of Figure~\ref{fig:CPD} compares the color--rotation period distribution of Theia 456 stars with stars from the Pleiades \citep[125 Myr according to][periods from \citeauthor{Rebull2016}~\citeyear{Rebull2016}]{stauffer1998} and Praesepe \citep[670 Myr;][]{Douglas2019}. 
Three features in the color--period distribution are  particularly important for estimating Theia 456's age. First is the existence of a coherent sequence of slowly rotating stars in the color range 0.5~$<$~$(BP - RP)$~$<$~1.25 mag and extending from periods of about a day to periods of about 10 days. The existence of such a sequence indicates that the stream's stars are indeed coeval, as period measurements for a group of unassociated field stars would not produce such a sequence.

Second is the location of this sequence between those of Pleiades and Praesepe members with the same colors. This suggests that Theia 456's age is between that of these two benchmark clusters. However, note that this comparison is dependent on accurate de-reddening for the stars in Theia 456. We have separately compared the stellar rotation periods against \teff\ for those stars observed by LAMOST and found that Theia 456 stars are closely overlapping with the Pleiades.

Third is the turnover of this slow-rotating sequence at a $(BP-RP)$ $\simeq1.3$. At an older age, such as that of Praesepe, that turnover occurs at redder colors and longer periods \citep[see also the comparison of the Pleiades with M34 by][]{Stauffer2016}. The similar positioning of the turnover between Theia 456 and the Pleiades suggests the two stellar populations have similar ages. 

Combined, our analysis from both isochrone fitting and gyrochronology suggests that the stars in Theia 456 have a common age of 150--200 Myr. A more precise measurement will require a more focused analysis, outside the scope of this work.

\subsection{Are There any Gradients across the Extent of Theia 456?}
\label{sec:gradient}

\begin{figure}
    \begin{center}
    \includegraphics[width=1.0\columnwidth]{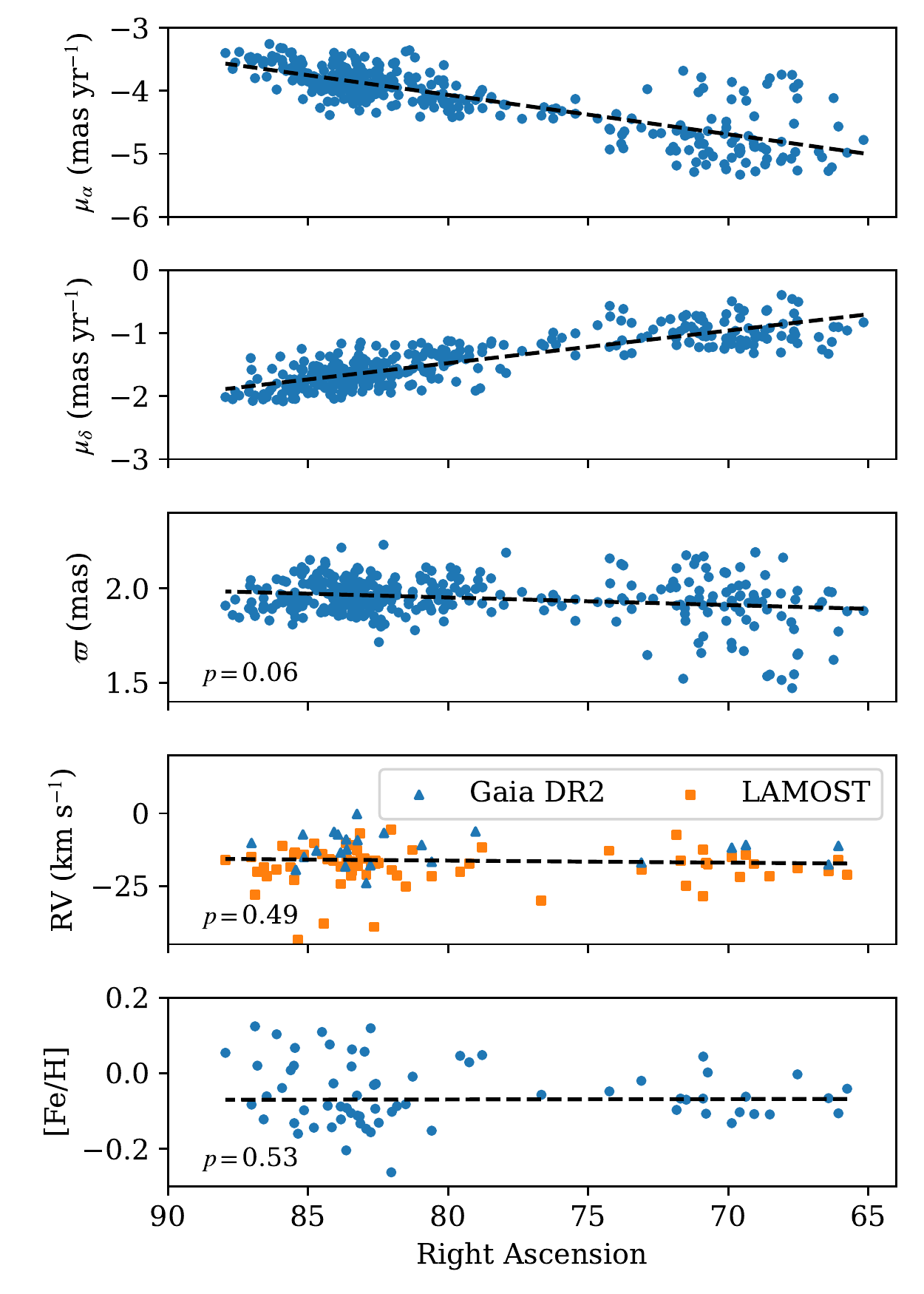}
    \caption{ We search for gradients in observables along the extent of Theia 456. Clear gradients exist in proper motion, as expected from geometric effects. $p$-values (calculated using Spearman's rank correlation) listed in the bottom left corner of the remaining three quantities (parallax, radial velocity, and [Fe/H]) indicate the null hypothesis cannot be rejected, and we therefore conclude that no significant correlations exist.} 
    \label{fig:gradient}
    \end{center}
\end{figure}

Finally, we search for the existence of any gradients in characteristics across the extent of Theia 456. In particular, Theia 456 appears in Figure \ref{fig:Fe_H} to have two wings, one more dense toward the Southeast and one less dense toward the Northwest, with a bridge connecting the two. Using the right ascension as a metric for each member star's position across the extent of Theia 456, Figure \ref{fig:gradient} shows how each observable does, or does not, vary. Since Theia 456 spans nearly 20$^{\circ}$ across the sky, geometric effects cause even two stars moving together to have different proper motions. Therefore, the gradients in $\mu_{\alpha}$ and $\mu_{\delta}$ in the top two panels of Figure \ref{fig:gradient} are expected; we analyze the dynamical history of Theia 456 in Section \ref{sec:dynamics}. 

The third panel of Figure \ref{fig:gradient} shows a slight gradient in the parallax across the extent of Theia 456, suggesting the structure may lie at an angle to the Sun, rather than perpendicular. However, a Spearman's $\rho$ correlation coefficient assigns only a marginal significance to this gradient, with a $p$-value of 0.06. Therefore, we cannot reject the null hypothesis that no correlation exists. A more dedicated analysis of the membership of Theia 456 will either confirm this gradient or alternatively could find that the slight observed gradient is due to some contamination in the sample at low right ascensions. In the bottom two panels of Figure \ref{fig:gradient}, we show the radial velocities and [Fe/H] measurements as a function of right ascension. $p$-values listed in the bottom left corner of each of these panels indicates the lack of a gradient in either of these quantities. We therefore conclude, at least with our current Theia 456 membership catalog and measurement precision, the chemical characteristics of Theia 456 are consistent across its entire extent.

\section{Dynamical Origin}
\label{sec:dynamics}

\begin{figure}
    \begin{center}
    \includegraphics[width=1.0\columnwidth]{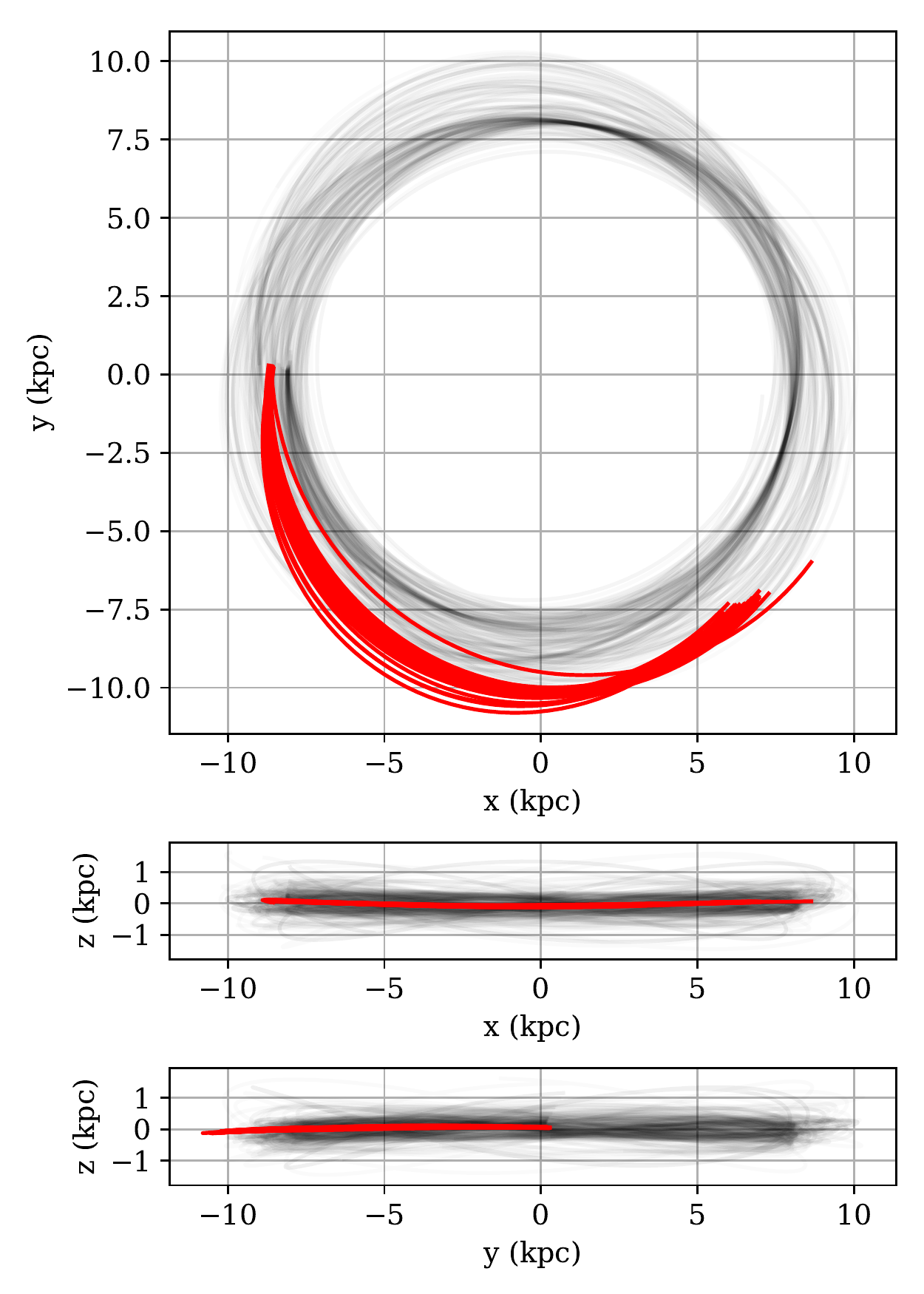}
    \caption{ We follow the orbit for 100 Myr backwards in time for the Theia 456 stars with radial velocities measured by \gaia\ (red). Compare with the orbits of 100 random Gaia DR2 disk stars (black). With its low $z$-height and low eccentricity, Theia 456 is part of the Milky Way thin disk population. }
    \label{fig:orbit}
    \end{center}
\end{figure}

\begin{figure*}
    \begin{center}
    \includegraphics[width=1.0\textwidth]{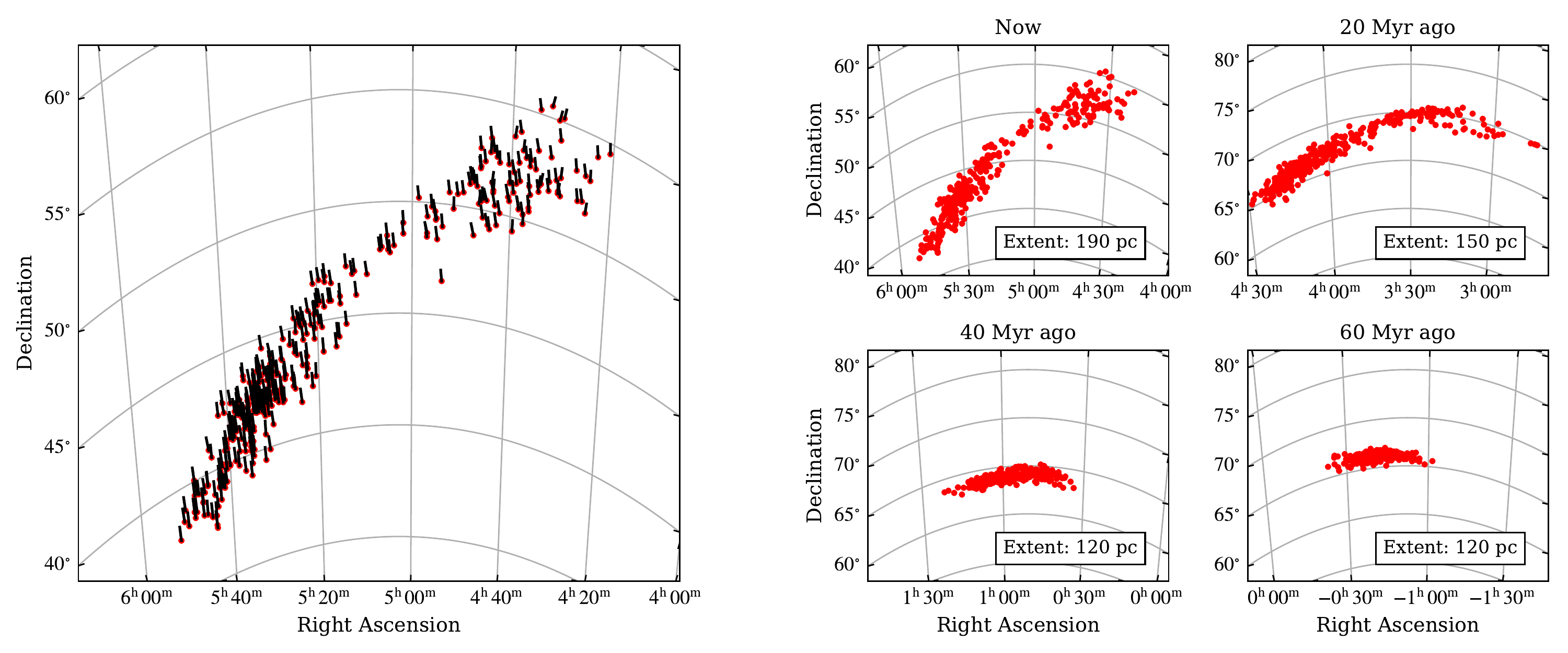}
    \caption{ The current positions of Theia 456 members, with lines indicating their proper motions (left panel). The positions of stars are integrated backwards in a Milky Way potential, assuming all stars have a radial velocity of $-$12.09 km s$^{-1}$, then mapped onto the sky, accounting for the Sun's motion around the Milky Way. The right four panels show that the structure appears more compact further back in time. Although some of this effect is due to the fact that Theia 456 was further away from the Sun in the past, we list the spatial extent of Theia 456 in the bottom right of each panel. When integrating backwards, the two ends of Theia 456 appear to have a common origin in a more compact structure. }  
    \label{fig:dynamics}
    \end{center}
\end{figure*}

\gaia\ astrometry provides five of the six dimensions of phase space for stars in Theia 456. With the addition of radial velocities from either \gaia\ DR2 or LAMOST, we can track the motion of a subset of Theia 456 stars backward in time. We use the python package {\tt gala} \citep{gala} to integrate the motion of these stars around the Milky Way's gravitational potential. Integrations are performed within the built-in {\tt MilkyWayPotential} potential using {\tt Astropy v4.0} defaults \citep{astropy1, astropy}, which combines an NFW halo potential \citep{NFW} with the disk model from \citet{galpy}. Figure \ref{fig:orbit} shows the backwards trajectory of these stars (red lines) over the past 100 Myr. Compared with randomly chosen \gaia\ disk stars (gray lines), the top panel (face-on) and bottom two panels (edge-on, from each side) shows that Theia 456 is clearly part of the Milky Way's thin disk.

Compared with the positions of Theia 456 stars today (near $y=0$), the positions our calculations derive for these stars 100 Myr in the past are distributed over several kpc. One may be tempted to conclude that these stars lack a common dynamical origin; however, we find the spread in the ending positions of our orbital integration is due to measurement imprecision. For instance, an imprecision of 1 km s$^{-1}$ in the radial velocity translates to a spread of 100 pc in distance when traveling in a flat gravitational potential. When placed in a Milky Way potential, the combination of these measurement imprecisions and parallactic differences can easily lead to $\sim$kpc differences in the calculated positions after 100 Myr. If one wants to exactly calculate the dynamical evolution of the stars in Theia 456, back to their birth 150--200 Myr ago, more precise radial velocities and parallaxes are required. Of course, an accurate model of both the Milky Way potential and the Sun's phase-space position is required to calculate the exact birth location of Theia 456 stars. Furthermore, clumpiness in the Milky Way potential due to spiral arms, the Milky Way bar, and passing globular clusters and giant molecular clouds is extremely difficult to model and can potentially have a profound impact on the structure of stellar groups \citep{gieles2006, gustafsson2016, price-whelan2016, kamdar2021}. However, we are more interested in the shape and spread of the distribution of Theia 456 stars at formation, which is somewhat less sensitive to Milky Way substructure, rather than the exact birth locations of Theia 456 stars.

Nevertheless, we can approximate the backward evolution by repeating the orbital integration while adopting a constant radial velocity of $-11.5$\,km\,s$^{-1}$ for each star in Theia 456 based on the median \gaia{} DR2 radial velocity of the stars where it is measured. This approximation allows us to estimate the prior dynamical evolution of all Theia 456 stars, not just of those with measured radial velocities. The left panel of Figure \ref{fig:dynamics} shows the sky positions of all Theia 456 stars today, with attached lines indicating their proper motion vectors backward in time. In the right four panels, we progressively integrate the orbits further back in time to show the how the sky positions of Theia 456 stars changed; we find that 60 Myr ago the stars in Theia 456 contracted into a more compact structure on the sky. At every time step, we additionally calculate the physical extent of the structure, listed in the bottom right of the four righthand panels. One can see from the sky positions of these stars that the two wings to Theia 456 move closer together, a trend confirmed by the smaller size of the cluster in the past. Note that the compactness we observe is dependent upon our adoption of a constant radial velocity for all stars in Theia 456, a significant simplifying approximation. An improved dynamical analysis will require radial velocities with precision $<$1\,km\,s$^{-1}$ \citep{ducourant2014, donaldson2016, crundall2019}.

\section{Discussion and Conclusions}
\label{sec:conclusions}

We have carefully analyzed the membership, kinematics, metallicity, rotation periods, and dynamical origin of Theia 456, one of the several thousand Theia structures identified by \citet{kounkel2019} and \citet{kounkel2020} in \gaia\ DR2. We produce a more robust membership catalog of 362 stars. We further confirm the kinematic coherence of these stars, analyzing both the astrometric parameters provided by \gaia\ and radial velocities provided by both \gaia\ and LAMOST. Using [Fe/H] values measured by LAMOST, we find that Theia 456 stars have similar, slightly sub-solar iron abundances of [Fe/H]$=-0.07\pm0.12$ dex, with the scatter dominated by systematics in the LAMOST [Fe/H] measurements. After carefully analyzing the [Fe/H] data, we conclude that Theia 456 stars have a common metallicity.

We additionally cross-match the Theia 456 catalog with ZTF and TESS photometric catalogs, allowing us to find photometric rotation periods for 216 stars. Comparing these to other known stellar populations with well-measured rotation periods and well-constrained ages, we estimate that Theia 456 has an age of 150--200 Myr. This is consistent with the 165$^{+70}_{-50}$ Myr age \citet{kounkel2019} derive using machine learning methods. Differential extinction prevents a more precise estimate through photometry alone. 

Taken together, these results strongly suggest that the stars in Theia 456 have a common origin, forming at around the same time from chemically well-mixed material. Is this the result of the monolithic collapse paradigm for star formation \citep{lada2003} in which gas clouds collapse and then are tidally disrupted over hundreds of Myr? Our analysis integrating the orbits of Theia 456 stars within a Milky Way potential suggests it originated in a more compact structure. However, our astrometric measurements are not precise enough to differentiate between whether that structure initially began as a more spherical shape as would have been produced by the monolithic collapse of a molecular cloud or if Theia 456 stars formed out of the hierarchical collapse of a more extended structure \citep{elmegreen2002, elmegreen2008}. Many of the Theia objects connect known open clusters with larger, low-density stellar structures  surrounding them \citep{kounkel2019, kounkel2020}, which could be consistent with both theories, depending on whether those stellar streams are the remnants of tidal stripping of the open clusters or whether the open clusters are simply the most overdense regions of an extended, hierarchically collapsing, filamentary cloud. 

Theia 456 itself appears to contain two separate regions, one in the Southeast and one in the Northwest, connected by a bridge. We find no significant difference in the bulk kinematic or metallicity characteristics between the two regions, and our dynamical analysis, summarized by Figure \ref{fig:dynamics}, indicates the two regions appear to coalesce in the past. While a further, more detailed analysis of stellar membership is required to confirm the bimodal structure of Theia 456, we speculate that the mere existence of two separate cores lends credence to a hierarchical formation scenario. A dedicated dynamical analysis, which is outside the scope of this work, is required to confirm that conclusion.

Our analysis of Theia 456 has focused on its origin; the corollary is to ask what will Theia 456 look like in 100 Myr. Our astrometric measurements currently lack the precision to accurately integrate the orbits of these stars too far into the future. Nevertheless, we can predict that the combination of the Galactic tide, dynamical motions internal to Theia 456, and passing Milky Way substructures---both dark matter and baryonic---will progressively disrupt the coherency of Theia 456. It seems inevitable that at some point in the future, possibly only $\sim$100 Myr hence, Theia 456 will be undiscoverable. We therefore speculate that some fraction of the disk stars comprising the Milky Way, perhaps even the Sun itself, formed in a loosely associated structure like that of Theia 456. Future \gaia\ data releases, with their improved parallax and proper motion measurements will aid with both confirming the nature of Theia structures as well as identifying new, as-yet undiscovered associations.

While our analysis suggests that the majority of stars identified as Theia 456 share a common origin, follow-up observations providing detailed chemical abundances can be used to further confirm its co-chemical and coeval nature. If, in fact, such analysis confirms that the stars in this and other Theia objects have a common origin, these will become essential to addressing unsolved problems ranging from planet formation and evolution to the dependence on age of fundamental stellar properties. These questions are best addressed by studying single-age stellar populations, but we have been limited by the relatively small number of known, accessible stellar structures in the solar neighborhood. Theia 456 is just one example of how \gaia\ is transforming stellar astrophysics.

\acknowledgements
We thank the anonymous referee for their careful reading of the manuscript and their insightful suggestions.

We thank Soichiro Hattori for sharing an early version of his TESS causal pixel modeling code, \texttt{tess\_cpm}, prior to its first official release as \texttt{unpopular}  \citep[][]{Hattori2021_tesscpm}. 
We measured periods from the resulting light curves using an interactive tool built by high school students participating in the Science Research Mentoring Program at the American Museum of Natural History: we thank Angeli Pante, Isabella Fraczek, and Linus Brooks for their contribution to that tool.

J.J.A.\ acknowledges support from CIERA and Northwestern University through a Postdoctoral Fellowship.
J.L.C.~and M.A.A.~acknowledge support for this work from the TESS Guest Investigator program under NASA grant 80NSSC22K0299. J.L.C.~acknowledges support provided by the NSF through grant AST-2009840.
J.C.~acknowledges support from the Agencia Nacional de Investigación y Desarrollo (ANID) via
Proyecto Fondecyt Regular 1191366; and from ``Centro de Astronomía y
Tecnologías Afines" project BASAL AFB-170002. S.C.S.~was supported by a Research Innovation and Scholarly Excellence (RISE) grant from the University of Tampa.

This work has made use of data from the European Space Agency (ESA) mission {\it Gaia},\footnote{\url{https://www.cosmos.esa.int/gaia}} processed by the {\it Gaia} Data Processing and Analysis Consortium (DPAC).\footnote{\url{https://www.cosmos.esa.int/web/gaia/dpac/consortium}} Funding for the DPAC has been provided by national institutions, in particular the institutions participating in the {\it Gaia} Multilateral Agreement.

Guoshoujing Telescope (the Large Sky Area Multi-Object Fiber Spectroscopic Telescope LAMOST) is a National Major Scientific Project built by the Chinese Academy of Sciences. Funding for the project has been provided by the National Development and Reform Commission. LAMOST is operated and managed by the National Astronomical Observatories, Chinese Academy of Sciences.

\software{  {\tt astropy} \citep{astropy}, 
            {\tt astroquery} \citep{astroquery}, 
            {\tt gala} \citep{gala_zenodo}, 
            {\tt NumPy} \citep{numpy}, 
            {\tt SciPy} \citep{scipy}, 
            {\tt matplotlib} \citep{matplotlib}, 
            {\tt scikit-learn} \citep{scikit-learn}, 
            \texttt{TESScut} \citep{Astrocut},
            {\tt unpopular} \citep{Hattori2021_tesscpm}, 
            The IDL Astronomy User's Library \citep{IDLastro}
            }
  
\appendix

\section{Catalog of Theia 456 Members} \label{b:catalog}

We provide a catalog of the 362 members we identify within Theia 456. Table \ref{tab:catalog} provides a sample of this data, which includes only a subset of the columns for the first 10 members. The complete dataset, which is publicly available, contains all the data from \gaia\ EDR3, as well as LAMOST data, our derived extinctions for the three \gaia\ photometric bands, and rotation periods from ZTF and TESS.

\begin{table*}
  \centering
  \caption{Theia 456 Catalog}
  \label{tab:catalog}
  \begin{tabular}{lcccccccccccc}
  \hline
   & \multicolumn{4}{c}{\gaia} & & \multicolumn{3}{c}{LAMOST} & & ZTF & & TESS \\
   \cline{2-5}\cline{7-9}\cline{11-11} \cline{13-13}
  \gaia\ ID & RV & $G$ & $BP$ & $RP$ & & RV & [Fe/H] & $T_{\rm eff}$ & & 
  $P_{\rm rot}$ & & $P_{\rm rot}$ \\
   & (km s$^{-1}$) & (mag) & (mag) & (mag) & & (km s$^{-1}$) & & ($K$) & & (days) &  & (days) \\
  \hline
    191263006888303360 & $\cdots$ & 16.64 & 17.59 & 15.65 & & -37.9$\pm$6.14 & -0.652$\pm$0.158 & 3911$\pm$166 && $\cdots$ && $\cdots$ \\
    208504792317742720 & $\cdots$ & 15.03 & 15.56 & 14.29 && $\cdots$ & $\cdots$ & $\cdots$ && 8.53 && 8.13 \\
    257811119959501952 & $\cdots$ & 14.69 & 15.29 & 13.94 && $\cdots$ & $\cdots$ & $\cdots$ && $\cdots$ && -14.06 \\
    256545856951806720 & $\cdots$ & 16.18 & 17.00 & 15.30 && $\cdots$ & $\cdots$ & $\cdots$ && 9.57 && $\cdots$ \\
    256511011879223296 & $\cdots$ & 15.13 & 15.70 & 14.41 && $\cdots$ & $\cdots$ & $\cdots$ && 7.80 && $\cdots$ \\
    256436696063404416 & $\cdots$ & 16.66 & 17.48 & 15.65 && $\cdots$ & $\cdots$ & $\cdots$ && 3.87 && $\cdots$ \\
    255788396520289792 & $\cdots$ & 14.76 & 15.30 & 14.06 && $\cdots$ & $\cdots$ & $\cdots$ && 6.75 && $\cdots$ \\ 
    211934199151015936 & $\cdots$ & 17.34 & 18.51 & 16.29 && $\cdots$ & $\cdots$ & $\cdots$ && 2.06 && $\cdots$ \\
    208725416200937856 & $\cdots$ & 17.50 & 18.53 & 16.42 && $\cdots$ & $\cdots$ & $\cdots$ && $\cdots$ && $\cdots$ \\
    207653869102503808 & $\cdots$ & 17.83 & 19.11 & 16.74 && $\cdots$ & $\cdots$ & $\cdots$ && $\cdots$ && $\cdots$ \\
  \hline
  \end{tabular}
\end{table*}

\section{Details of the Rotation Period Analysis} \label{a:rotation}

We measured rotation periods using time series imaging data acquired by the space-based TESS and ground-based ZTF facilities. For TESS, we downloaded 40$\times$40 pixel image cutouts with \texttt{TESScut} \citep{Astrocut}, then extracted light curves using the TESS causal pixel modeling package \texttt{unpopular} \citep{Hattori2021_tesscpm}. Briefly, this tool uses pixels outside of an exclusion region centered on the target (the central 6$\times$6 pixels in our implementation) to model and subtract off the time-varying systematics (e.g., Earthshine reflected into the TESS optics). Next, we calculate Lomb--Scargle periodograms \citep{Lomb1976, Scargle1982, PressRybicki1989} with \texttt{astropy.timeseries} with test periods spaced logarithmically between 0.1 and 30 days. We visually inspected the light curves, periodograms, and phase-folded light curves to select the optimal sectors or segments of data to use, identify cases where the periodogram preferred the 1/2-period harmonic period, and classify the light curve to validate the resulting period. 

For ZTF, following the procedure described by \citet{curtis2020}, we downloaded 8$'\times$8$'$ image cutouts from the NASA/IPAC Infrared Science Archive (IRSA). Next, we extracted light curves using simple aperture photometry for each target and a selection of neighboring stars queried from Gaia. The light curves for the target and references stars are normalized by subtracting off their median magnitudes, and then a systematics light curve (i.e., per epoch zeropoint corrections) is created by calculating the median magnitude of this cohort at each epoch to capture their systematically varying brightnesses. We subtract this systematic signal from the target's light curve to produce a refined light curve ready for period analysis. Our procedure typically improves the photometric precision over the light curves produced by the ZTF pipeline. We then apply a similar Lomb--Scargle approach to measure rotation periods as was applied to the TESS data.

% which will be  
% \textit{Accept}---we accept the period; 
% \textit{Q=B}---the period is added to our table but with a lower quality flag; 
% \textit{Unclear}---there is apparent variability, but the period is unclear;
% \textit{Flat}---no obvious variability present in the light curve; 
% \textit{Garbage}---the light curve itself is rejected, usually because the targ

% Jason's going to drop in all the details for the light curves and rotation period measurements here. This will include a figure set, one panel for each star with TESS and/or ZTF data, which will show what's available and the final Prot designation.

% \begin{itemize}
%     \item TESS: cpm, tesscheck, figure set
%     \item ZTF: curtis2020, etc, figure set
%     \item compare tess and ztf
%     \item more on gyro age.
% \end{itemize}

\bibliographystyle{aasjournal}
\bibliography{gaia}

\end{document}